\documentclass[twoside]{article}
\usepackage{fleqn,espcrc2,epsfig}

\usepackage{graphicx}

\newcommand{\AmS}{{\protect\the\textfont2
  A\kern-.1667em\lower.5ex\hbox{M}\kern-.125emS}}
\providecommand{\BBAR}{\mbox{$\overline{\mathrm B}$}}

\providecommand{\KS}{\mbox{${\mathrm K_{\mathrm S}^0}$}}

\providecommand{\VCB}{\mbox {$ \mathrm {V_{cb}}$ }}

\providecommand{\VUB}{\mbox {$ \mathrm {V_{ub}}$ }}

\providecommand{\VTB}{\mbox {$ \mathrm {V_{tb}}$ }}
\providecommand{\VCSSTAR}{\mbox {$ \mathrm {V_{cs}^*}$ }}

\providecommand{\VUBSTAR}{\mbox {$ \mathrm {V_{ub}^*}$ }}

\providecommand{\VTSSTAR}{\mbox {$ \mathrm {V_{ts}^*}$ }}

\providecommand{\BZ}{\mbox{${\mathrm B}^0_{\mathrm }$}}
\providecommand{\BZBAR}{\mbox{$\overline {{\mathrm B}^0_{\mathrm }} $}}

\providecommand{\BBBAR}{\mbox{ $ {\mathrm B} \overline {\mathrm B} $ }}

\providecommand{\DZ}{\mbox{${\mathrm D}^0$}}
\providecommand{\DZBAR}{\mbox{$\overline{{\mathrm D}^0}$}}

\providecommand{\DSTP}{\mbox{${\mathrm D}^{\ast+}$}}

\providecommand{\cleo}{\mbox{\sf CLEO}} 

\providecommand{\JPSI}{\mbox{$\mathrm{J}/\psi$}}

\providecommand{\UFOURS}{\mbox{$\Upsilon(4{\mathrm S})$}}

\providecommand{\EPEM}{\mbox{${\mathrm e}^+{\mathrm e}^-$}}
\providecommand{\epem}{\mbox{${\mathrm e}^+{\mathrm e}^-$}}

\providecommand{\B}{\mbox{${\mathrm B}$}}

\providecommand{\cleoii}{\mbox{\sf CLEOII}}
\providecommand{\cleoiiv}{\mbox{\sf CLEOII.V}}

\providecommand{\slowpi}{\mbox{$\pi_{\mathrm s}$}}

\providecommand{\cleoii}{\mbox{\sc CLEO \ II}}
\providecommand{\cleoiiv}{\mbox{\sc CLEO \ II.V}}

\providecommand{\etal}{\mbox{\it et al.}}

\providecommand{\BTOKSG}{\mbox{${\rm B} \rightarrow {\rm K}^{*}\gamma$}}

\providecommand{\BTOKSPG}{\mbox{${\rm B}^+ \rightarrow {\rm K}^{*+}\gamma$}}
\providecommand{\BTOKSZG}{\mbox{${\rm B}^0 \rightarrow {\rm K}^{*0}\gamma$}}

\providecommand{\KS}{\mbox{${\rm K}_{\rm s}^0$}}
\providecommand{\PZ}{\mbox{$\pi^0$}}

\def\CL95{6.5 \times 10^{-4}}

\def\Dzero{{\rm D}^0}

\def\DsDs{{\rm D}^{*+}{\rm D}^{*-}}
\def\B0DsDs{{\rm \overline B}^0\rightarrow\DsDs}

\def\slowpi{\pi^+_{\rm slow}}

\def\Dzero_kpipizero{{\rm D}^0\rightarrow{\rm K}^-\pi^+\pi^0}
\def\DzeroTo_K3pi{{\rm D}^0\rightarrow{\rm K}^-\pi^+\pi^+\pi^-}
\def\Dzeroto_Kspipi{{\rm D}^0\rightarrow{\rm K}^0_{\rm s}\pi^+\pi^-}
\def\DzerotoK0s2PIpizero{{\rm D}^0\rightarrow{\rm K}^0_{\rm s}\pi^+\pi^-{\pi^0}}

\providecommand\PSIP{\mbox{$\psi^{(\prime)}$}}

\renewcommand\BZ{\mbox{${\rm B}^0$}}
\renewcommand\BZBAR{\mbox{$\overline{{\rm B}^0}$}}
\renewcommand\KS{\mbox{${\rm K}^0_{\rm S}$}}

\providecommand\CPV{\mbox{$CPV$}}
\providecommand\ACP{\mbox{${\cal A}_{CP}$}}
\providecommand\BR{\mbox{${\cal B}$}}

\def\SAKSTP{\mbox{$+0.38\pm 0.20$}}  
\def\BAKSTP{\mbox{$+0.06\pm0.09$}}  

\def\SAKSTZ{\mbox{$-0.13\pm0.17$}} 
\def\BAKSTZ{\mbox{$-0.03\pm 0.08$}} 
 
\def\SAKST{\mbox{$+0.08\pm 0.13$}} 
\def\BAKST{\mbox{$+0.01\pm 0.06$}} 

\title{Search for $CP$ violation at \cleo}

\author{{David E. Jaffe\thanks{Now at Brookhaven National Laboratory, Upton, NY 11973}, University of California San Diego,
La Jolla, CA 92093}
       }
       
\begin{document}

\begin{abstract}
Recent results from \cleo\ on the search for $CP$ violation
in beauty and charm meson decays are reviewed.
\vspace{1pc}
\end{abstract}

\maketitle


 $CP$ violation (\CPV ) was first observed nearly 40 years
ago in the form of  mixing-induced \CPV\ in 
neutral kaon decays~\cite{ref:epsk}. With the 
 the recent confirmation of the observation of direct $CP$ violation
in neutral kaon decays~\cite{ref:epsprime}, only 
\CPV\ due to the interference between mixing and decay remains
to be observed. With the advent of the asymmetric B-factories,
this phenomenon may soon be observed with a 
measurement of $\sin 2\beta$~\cite{cdf,mh,ys}. The latter measurement
has the advantage of a nearly unambiguous interpretation 
in terms of the description of weak decays in the standard model (SM).

 \cleo\ has performed a number of 
searches for \CPV\  in beauty and charm
meson decays. By and large the asymmetries expected in the 
SM  are significantly smaller than the experimental
precision so the results are primarily searches for physics beyond
the SM.

 The \cleo\ results for B mesons are based upon 
$9.7\times 10^6$ \BBBAR\ pairs collected at the \UFOURS\ 
resonance with the \cleoii\ ($3.3\times 10^6$\BBBAR ) and \cleoiiv\ 
($6.4\times 10^6$ \BBBAR) detector configurations at the
CESR symmetric \EPEM\ colllider. The search for mixing and \CPV\ 
in neutral charm meson decays utilizes $9.0\ {\rm fb}^{-1}$ 
of \EPEM\ collisions at $\sqrt{s} \approx 10.6\ {\rm GeV}$
accumulated with the \cleoiiv\ configuration. The inner wire chamber
and 3.5 cm radius beampipe of \cleoii\ \cite{cleoii} were replaced by a 2.0 cm 
radius beampipe and a three-layer, double-sided silicon
vertex detector (SVX) to create \cleoiiv\ \cite{cleoiiv}. 
In addition the argon:ethane gas mixture
in the main drift chamber was replaced by a helium:propane mixture.
The resulting improvements in momentum and specific ionization
($dE/dx$) resolution permitted better separation of high momentum
($\sim 2.5\ {\rm GeV}/c$) charged kaons and pions. The SVX
also permits the measurement of the proper decay time of neutral
charm mesons that is essential for the study of \DZ\DZBAR\ mixing
phenomena.

 In the B system, \cleo\ has searched for evidence of direct \CPV\ 
through the measurements of rate asymmetries in charmless hadronic decays,
radiative decays and in $B^\pm\to\PSIP K^\pm$ decays.
Almost all  measurements rely on self-tagging decays with the charge
of a $K^\pm,\ \pi^\pm$ or $K^{*\pm}$ identifying the B or \BBAR\ at 
decay. The branching fractions of a number of charmless hadronic
decays observed by \cleo~\cite{cleo.rare,cinabro} are shown in 
Table~\ref{tab:charmless}.
 Table~\ref{tab:charmless} also contains the preliminary results of
the Belle~\cite{belle.rare} and BaBar~\cite{babar.rare}
experiments confirming the \cleo\ results.

\begin{table*}[htb]
\caption{Selected charmless hadronic B branching fractions in units of $10^{-6}$. All limits at 90\% CL. $\spadesuit$ = used for \cleo\ \CPV\  search. }
\label{tab:charmless}
\begin{tabular}{|c|c|c c c|}
\hline 
 & Final & \multicolumn{3}{c}{Experiment}\\
 & state & \cleo~\protect\cite{cleo.rare,cinabro} & BELLE~\protect\cite{belle.rare} & BABAR~\protect\cite{babar.rare}  \\
 \hline
$\spadesuit$&{$K^\pm\pi^\mp$}  &$17.2^{+2.5}_{-2.4}\pm1.2$ & $17.4^{+5.1}_{-4.6}\pm3.4$ &$12.5^{+3.0+1.3}_{-2.6-1.7}$ \\
$\spadesuit$ &{$K^0\pi^\pm$}  &$18.2^{+4.6}_{-4.0}\pm1.6$ & $16.6^{+9.8+2.2}_{-7.8-2.4}$ & {}\\
$\spadesuit$ &{$K^\pm\pi^0$}  &$11.6^{+3.0+1.4}_{-2.7-1.3}$ & $18.8^{+5.5}_{-4.9}\pm2.3$ & {}\\
&{$K^0\pi^0$}  &$14.6^{+5.9+2.4}_{-5.1-3.3}$ & $21.0^{+9.3+2.5}_{-7.8-2.3}$ & {}\\
\hline
&{$\pi^\pm\pi^\mp$}  &$4.3^{+1.6}_{-1.4}\pm0.5$ & $6.3^{+3.9}_{-3.5}\pm1.6$ &$9.3^{+2.6+1.2}_{-2.3-1.4}$ \\
&{$\pi^\pm\pi^0$}  & $<12.7$ & $<10.1$ & \\
&{$\pi^0\pi^0$}  & $<5.7$ &  & \\
\hline
$\spadesuit$&$\eta'K^\pm$&$80^{+10}_{-9}\pm7$& &$62\pm18\pm8$\\
$\spadesuit$&$\omega \pi^\pm$&$11.3^{+3.3}_{-2.9}\pm 1.4$&&\\
&$\phi K^\pm$&$6.4^{+2.5+0.5}_{-2.1-2.0}$&$17.2^{+6.7}_{-5.4}\pm 1.8$& \\
\hline
\end{tabular}
\end{table*}

 In the SM charmless hadronic B meson decays occur through
$b\to u$ (``tree'') or $b\to s$ (``penguin'') transitions.
The relatively large rate of $B\to K\pi$ with respect to $B\to \pi\pi$ 
indicates that the amplitudes for the tree ($A_T$) and penguin 
($A_P$) contributions are comparable. Interference between the 
$b\to u$ and $b\to s$ processes make both the branching fractions
($\propto |A_P/A_T|\cos\gamma\cos\delta$) and rate asymmetries 
($\propto |A_P/A_T|\sin\gamma\sin\delta$) sensitive to the
weak mixing angle $\gamma \approx \arg(-\VUBSTAR)$. The non-\CPV\ 
phase difference is $\delta$ and is frequently referred to as the 
``strong'' phase. Based on the relative $B\to K\pi$ and $B\to \pi\pi$
rates, we have $|A_P/A_T|\sim 1/4$ while measurements of 
$|\VCB|$, $|\VUB|$, $\Delta m_d$, and $\epsilon_K$ indicate that
$\gamma \sim 90^\circ$. Thus a large strong phase $|\sin\delta| \sim 1$
could produce rate asymmetries of ${\cal O}(50\%)$ that would be
observable with the current \cleo\ data.

\begin{figure}[htb]
\epsfig{figure=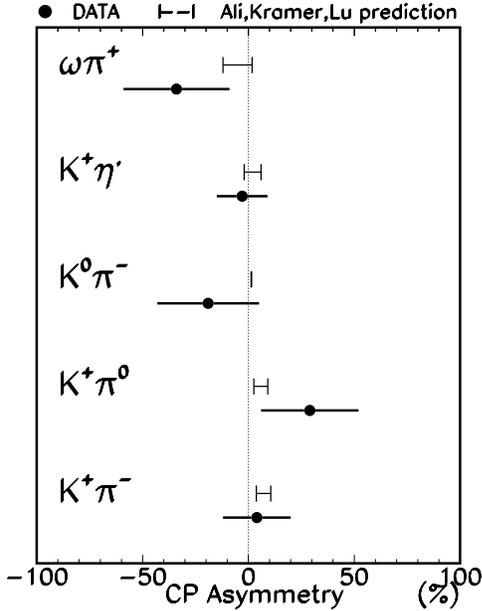,width=\linewidth}
\caption{\cleo\ results~\protect\cite{cleo.acp.rare} for
the charge asymmetry for five charmless hadronic B meson decays.
The prediction~\protect\cite{akl} 
assumes factorization, no soft final state interactions, $\rho = 0.12$ and $ \eta = 0.34$.}
\label{fig:acp.rare}
\end{figure}

\cleo\ utilizes the unbinned maximum likelihood (ML) method to achieve
maximum precision on the charmless hadronic branching fractions.
The ML technique utilizes the observables
$\Delta E \equiv E(B) - E_{\rm beam}$ and 
$M^2(B) \equiv E^2_{\rm beam} - {\bf p}^2(B)$ where
$E(B)$ and $ {\bf p}(B)$ are the energy and momentum of the 
B candidate, respectively, $dE/dx$, the masses of intermediate resonances
and the helicity angle of $B\to {\rm vector},{\rm pseudoscalar}$ decays
where applicable. In addition event shape variables are combined in
a Fisher discriminant that maximizes the separation between the 
``jetty'' $\EPEM\to q\bar{q}$ ($q = u,c,s,d$) background and the more
spherical \BBBAR\ decays. The likelihood is simultaneously maximized for
the branching fraction 
$\BR \equiv \frac{1}{2}(\BR(\BBAR\to \bar{f}) + (\BR({\rm B}\to {f}))$
and asymmetry $\ACP \equiv (\BR(\BBAR\to \bar{f}) - (\BR({\rm B}\to {f}))/(\BR(\BBAR\to \bar{f}) + (\BR({\rm B}\to {f}))$ 
to obtain the results~\cite{cleo.acp.rare} for the
five decay modes shown in Figure~\ref{fig:acp.rare}. All measured \ACP\ 
are consistent with zero and with the prediction shown indicating that the
strong phases are small for these decays. The precision of the measurements
varies between 10\% and 25\% and is entirely dominated by statistics. 
Systematic checks show that no artificial asymmetries are introduced by
either momentum or $dE/dx$ measurements at less than 1\% based on studies
of kinematically identified $K^\pm$ and $\pi^\pm$ from D decays.

 Radiative B meson decays, in contrast to the charmless hadronic decays,
are dominated by $b\to s\gamma$ transitions in the SM. This situation
is quantified by the good agreement between the measured inclusive
rate $\BR(b\to s\gamma)$~\cite{cleo.bsg.incl} and the 
next-to-leading order calculation~\cite{theory.bsg.incl} as shown in 
Table~\ref{tab:bsg}. Despite this agreement it is possible that
non-SM propagators could produce significant asymmetry ${\cal O}(40\%)$ 
in both inclusive and exclusive radiative B decays~\cite{theory.radb.acp}.

\begin{figure}[htb]
\epsfig{figure=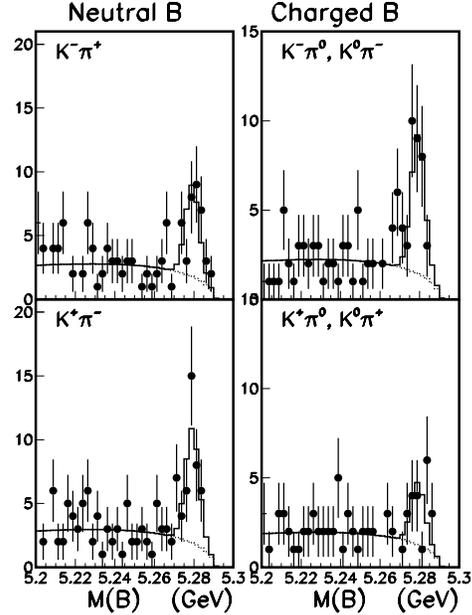,width=\linewidth}
\caption{The fitted $M(B)$ distributions for neutral and charged 
B and \BBAR\ candidates for \BTOKSG\ decays.}
\label{fig:acp.bsg}
\end{figure}

 The search for \CPV\ in \BTOKSG\ decays utilizes the self-tagging
\BTOKSPG\ ($K^{*+}\to K^0\pi^+, K^+\pi^0$) and \BTOKSZG\ ($K^{*0}\to K^+\pi^-$)
decays.
Only $\sim 60\%$ of the \BTOKSZG\ candidates are amenable to self-tagging
because the kinematic and $dE/dx$ 
identification of $K^{*0}$ and $\bar{K}^{*0}$ is 
ambiguous when $|{\bf p}_K| \approx |{\bf p}_\pi|$.
Suppression of backgrounds from
$\epem\to q\bar{q}\gamma$ (initial state radiation) and 
$\epem\to \pi^0 X$ is accomplished by requirements on the angle
of the $\gamma$ with respect to the \epem\ collision axis $|\cos\theta|<0.71$ 
and by vetoing $\gamma$ consistent with a $\pi^0$ origin, 
respectively. Additional suppression of the jetty $q\bar{q}$ background 
is achieved by requirements on the angle between the $\gamma$ and the thrust
axis~\cite{thrust} of the rest of the event excluding the B candidate.
Asymmetries of 
 ${\cal A}_{CP} = \SAKSTZ$ and \SAKSTP\ 
 for the signal and \BAKSTZ\ and \BAKSTP\ for the background for 
neutral and charged \BTOKSG\ are determined from 
fits to the $M(B)$ distributions of 
B and \BBAR\ candidates shown in Figure~\ref{fig:acp.bsg}.
Assuming that \CPV\ would be independent of the light spectator
quark $\ACP(\BTOKSG) = \SAKST\ [{\BAKST}]$ for the signal [background]
where the uncertainty includes the systematic uncertainty 
of 2.5\%~\cite{cleo.bsg.excl}.

\begin{table*}[htb]
\caption{Measured exclusive and inclusive branching fractions ($\times 10^{-5}$) for radiative B meson decays.}
\label{tab:bsg}
\begin{tabular}{|c|cc|c|}
\hline
Expt &  \BTOKSZG\  & \BTOKSPG\ &$b\to s\gamma$\\
\hline
Theory~\protect\cite{theory.bsg.incl} & & & $32.8\pm 3.3$\\
CLEO~\protect\cite{cleo.bsg.excl,cleo.bsg.incl}  &  $4.55^{+0.72}_{-0.68} \pm 0.34$     & $3.76^{+0.89}_{-0.83}\pm 0.28$ &  $31.5\pm3.5\pm3.2\pm2.6$ \\
BELLE~\protect\cite{belle.bsg} &  $4.94\pm0.93^{+0.55}_{-0.52}$ & $2.81\pm1.20^{+0.55}_{-0.40}$ & $33.4\pm 5.0^{+3.4+2.6}_{-3.7-2.8}$\\
BABAR~\protect\cite{babar.bsg} & $5.4\pm0.8\pm0.5$ & &\\
\hline
\end{tabular}
\end{table*}

 The techniques used to measure the inclusive $b\to s\gamma$ 
branching fraction~\cite{cleo.bsg.incl,cleo.old.bsg.incl} 
have been adapted to measure
$\ACP(b\to s\gamma)$. The B flavor is determined either by 
detecting a charged lepton from the semileptonic decay of the
other B or by self-tagging through the ``pseudo-reconstruction'' of
$X_s (X_s = K\  {\rm and}\  \le4\pi)$ with $X_s\gamma$ kinematically 
consistent with $B\to X_s\gamma$. The mistag rate for 
 lepton tagging is  $0.112$  due almost entirely to \BZ\BZBAR\ mixing
while the mistag rate for the pseudo-reconstruction 
is either $0.082$ or $0.122$ depending on the amount and quality of the
particle identification information available. The preliminary 
measured asymmetry for the lepton tag (pseudo-reconstruction) 
is  $\ACP = +0.155\pm 0.147$ (  $\ACP = -0.152\pm 0.112$) where the 
uncertainty is statistical only. Studies revealed that asymmetries
in lepton, $K$ and $\pi$ identification and reconstruction are $<1\%$.
Multiplicative uncertainties due to continuum $\epem\to q\bar{q}$
and $b\to c$ background subtraction are $\sim 3\%$. The preliminary
combined result with all corrections applied is
$\ACP\ = (-0.063\pm 0.090[{\it s}] \pm 0.022[{\it a}])
        \times (1.00\pm 0.03[{\it m}])$
where {\it s}, {\it a} and {\it m} denote the statistical, 
additive systematic and
multiplicative systematic uncertainties, respectively,
or $-0.22<\ACP <+0.09$ at 90\% CL. This limit and the
results for exclusive radiative decays exclude a
significant fraction of the range allowed by non-SM processes but are still
far from the ${\cal O}(1\%)$ level predicted by the SM.

 The final search for direct \CPV\ is in $B^\pm\to\PSIP K^\pm$ decays
($\PSIP$ stands for \JPSI\ and $\psi({\rm 2S})$) that proceed by 
$b\to c\bar{c}s$. The direct \CPV\ asymmetry for these decays is
expected to be very small because the sub-dominant penguin process
($b\to sc\bar{c}$) is suppressed and has nearly the same weak
phase 
$\arg\left(\VCB\VCSSTAR/\VTB\VTSSTAR\right) \approx \lambda^2\eta + \pi$
($\lambda = 0.22, \eta \le 1$) as the dominant process.
 Non-SM effects could produce a
noticeable asymmetry if there is an appreciable strong phase difference
between the SM and non-SM amplitudes~\cite{theory.acp.psik}. The 
quark process $b\to c\bar{c}s$ is the same as that for the ``golden
mode''  $\BZ\to \JPSI\KS$ that is being used to measure $\sin 2\beta$.
An asymmetry in  $B^\pm\to\PSIP K^\pm$ decays, besides being evidence 
of non-SM physics, would indicate possible complications for the measurement
of $\sin 2\beta$ with  $\BZ\to \JPSI\KS$.

\begin{figure}[htb]
\epsfig{figure=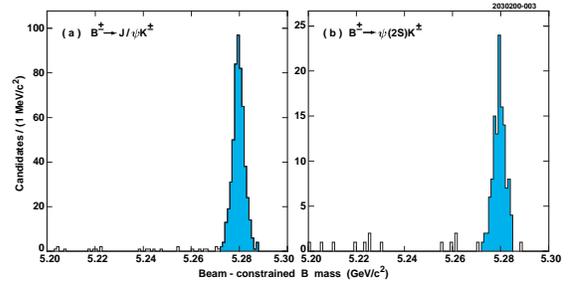,width=1.0\linewidth}
\caption{The M(B) distribution of (a) $ B^\pm\to\JPSI K^\pm$ and (b) $ B^\pm\to\psi(2{\rm S}) K^\pm$
 candidates. The shading indicates the candidates selected for the
asymmetry measurement.}
\label{fig:psik}
\end{figure}

 Experimentally  $B^\pm\to\PSIP K^\pm$ is nearly as background-free
as $\BZ\to \PSIP\KS$. The \PSIP\ are reconstructed in the
$\PSIP\to\ell^+\ell^-$ ($\ell=e,\mu$) and $\psi(2{\rm S})\to \JPSI\pi^+\pi^-$
modes. The charged kaon is identified kinematically to avoid any possible
$dE/dx$-induced bias and  $B^\pm\to\PSIP K^\pm$ candidates are selected
by requiring $|\Delta E/\sigma(\Delta E)| < 3$ and 
$|M(B)-M_{B^+}|/\sigma(M(B)) < 3$ as shown in Figure~\ref{fig:psik} 
where $\sigma(x)$ is the candidate-by-candidate uncertainty in $x$
as calculated from the covariance matrices of the reconstructed
charged tracks. A small correction of $(+0.3\pm0.3)\%$ is applied to the 
measured asymmetry to take into account the different cross-sections of
$K^+$ and $K^-$ in the \cleo\ detector material. 
The asymmetries $\ACP(J/\psi K^\pm) =   (+1.8\pm 4.3\pm 0.4)\%$
and $\ACP(\psi(2{\rm S})K^\pm) =  (+2.0\pm9.1\pm1.0)\%$ are consistent
with zero and are currently the most precise measurements of direct
\CPV\ in B meson decays~\cite{cleo.acp.psik}.

In contrast to B decays, the \cleoiiv\ SVX permits measurement of
the proper time dependence of charm meson decays~\cite{cleo.charm.life}
and enables the search for \CPV\ and \DZ\DZBAR\ mixing. \DZ\DZBAR\ mixing
is thought to be both GIM- and Cabibbo-suppressed
in the SM although a wide range of predictions exists~\cite{harry.compilation}
and recent re-evaluations indicate that the suppression may be
only ${\cal O}(0.1\%)$~\cite{bigi,petrov}. \DZ\DZBAR\ mixing through either
virtual or real intermediate states is quantified by the 
dimensionless parameters $x\equiv\Delta m/\Gamma$ and $y\equiv\Delta\Gamma/2\Gamma$, respectively, where $\Delta m$ and $\Delta \Gamma$ are the
mass and width differences of the mass eigenstates and $1/\Gamma$ 
is the average of the \DZ\ and \DZBAR\ lifetimes. 
Non-SM effects could produce
such signatures as $|x|\gg|y|$ and/or large ${\it Im}(x)/x$ (\CPV).
\cleo\ has searched for \DZ\DZBAR\ mixing by comparing the rate of
the ``wrong sign'' (WS) process $\DZ\to K^+\pi^-$ with that of
the ``right sign'' (RS) $\DZ\to K^-\pi^+$ decay where the initial
\DZ\ is identified by the charge of the pion in the strong decay
$\DSTP\to \DZ\slowpi$. 
For $|x|\ll 1$ and $|y|\ll 1$,
the proper time dependence of the WS rate is 
\begin{equation}
\label{eqn:ws}
r_{\rm ws}(t) 
= ( R_{\rm D} 
+ \sqrt{R_{\rm D}}{y'}t
+\frac{1}{4}({x'}^2 + {y'}^2)t^2) e^{-t}
\end{equation}
\noindent in units of the \DZ\ lifetime where $R_{\rm D}$ is the doubly-Cabibbo suppressed (DCS) rate, 
$y'\equiv x\sin\delta - y\cos\delta$, 
 $x'\equiv x\cos\delta + y\sin\delta$, and $\delta$ is a possible strong
phase between the DCS and mixing amplitudes. The observation of a significant
quadratic dependence in the proper time dependence of the WS rate would
be an indication of mixing through $x'$ or $y'$ while a linear dependence
would indicate mixing through $y'$.

\begin{figure}[htb]
	\epsfig{figure=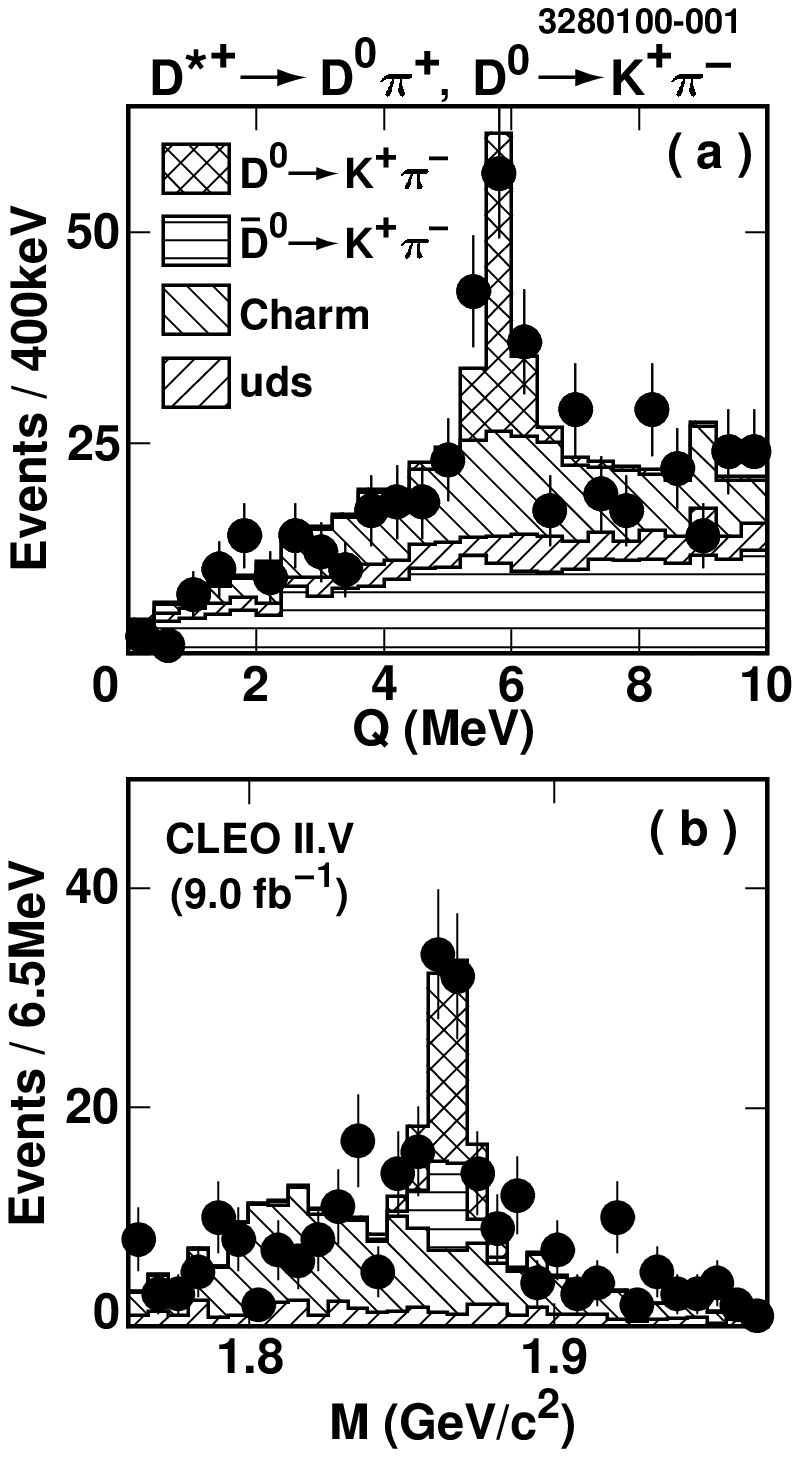,width=1.\linewidth}
\caption{The $Q$ and $M$ data and fit projections for the $\DZ\to K^+\pi^-$ candidates.}
\label{fig:kpi.ws}
\end{figure}

\begin{figure}[htb]
	\epsfig{figure=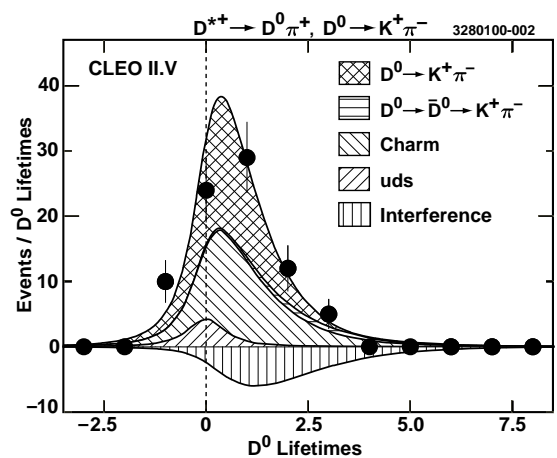,width=\linewidth}
\caption{ The fitted proper time distribution of  $\DZ\to K^+\pi^-$ candidates.}
\label{fig:kpi.t}
\end{figure}

 The WS rate is determined from a binned ML fit to the distribution of
WS candidates in the $Q$ {\it vs} $M$ plane ($M\equiv M(K\pi)$, 
$Q\equiv M(K\pi\slowpi) - M - M_{\pi^+}$). The shapes of the four distinct
backgrounds $\epem\to q\bar{q} \ (q=u,s,d)$, $q\bar{q}\to c\bar{c}$, 
$\DZ\to {\rm pseudoscalar},{\rm vector}$ and $\DZBAR\to K^+\pi^-$ are
taken from a simulated event sample corresponding to ten times the data
luminosity. The signal shape is taken from the RS data which has measured
resolutions of $\sigma(Q) = 190 \pm 6 \ {\rm keV}$ and 
$\sigma(M) = 6.4 \pm 0.1\ {\rm MeV}$. The superb $Q$ resolution is possible
due to the SVX and is achieved by fitting the $\slowpi$\ to the \DSTP\ 
production point that is taken as the intersection of the beam spot
and \DZ\ pseudotrack. A clear signal is visible in 
Figure~\ref{fig:kpi.ws} that shows the $Q$ and $M$ projections of the
WS candidates  when $M$ and $Q$ are required to be within $2\sigma$
of the known \DZ\ mass and \DSTP\ energy release, respectively.
The proper time distribution of the WS candidates within $2 \sigma$
of the RS signal region in $M$ and $Q$ is shown in Figure~\ref{fig:kpi.t}
along with a fit incorporating Eqn.~\ref{eqn:ws} with the modifications
$R_{\rm D}\to R_{\rm D}(1\pm A_{D})$, 
$x'[y']\to x'[y'](1\pm A_{M})^{\frac{1}{2}}$ and 
$\delta\to\delta\pm \phi$ where $+$($-$) corresponds to \DZ(\DZBAR)
for direct \CPV, mixing-induced \CPV\ and \CPV\ due to the interference
between mixing and decay, respectively. The fit prefers
$y' < 0$ (destructive interference) but the mixing parameters
$y'$ and $x'$ as well as the three $CP$ violating parameters are
all consistent with zero at 95\% CL (Table~\ref{tab:kpi})~\cite{cleo.dmix}.

\begin{table*}[htb]
\caption{The results of the most general fit to the $\DZ\to K^+\pi^-$
proper time distribution.}
\label{tab:kpi}
\begin{tabular}{ccc}
\textbf{} & \textbf{Central value} & \textbf{95\% C.L.}\\ \hline
\vphantom{$A^{A^A}$}$R_D$      & $(0.48\pm0.12\pm0.04)\%$ & 
$0.24\%<R_D<0.71\%$\\
$y^\prime$ & $(-2.5^{+1.4}_{-1.6}\pm0.3)\%$ & $-5.8\%<y^\prime<1.0\%$ \\ 
$x^\prime$ & $(0\pm1.5\pm0.2)\%$             & $|x^\prime|<2.9\%$ \\
$(1/2)x^{\prime2}$ &     & $<0.041\%$ \\ 
 & \multicolumn{2}{c}{CP violating parameters} \\
$A_M$ & $0.23^{+0.63}_{-0.80}\pm 0.01$ & No Limit \\
$A_D$ & $-0.01^{+0.16}_{-0.17}\pm 0.01$ & $-0.36<A_D<0.30$\\
$\sin \phi$ & $0.00 \pm 0.60\pm 0.01$ & No Limit \\ \hline 
\end{tabular}
\end{table*}

 Preliminary results of a similar analysis
for the WS process $\DZ\to K^+\pi^-\PZ$ reveal an
excess of  $N_{WS}=39{}^{+10}_{-9}\ \pm7$\ candidates~\cite{cinabro}.
Lack of knowledge of the WS resonant substructure (Dalitz plot)
confounds the interpretation of this preliminary observation both for the
relative WS to RS rate and for the proper time dependence.
In essence each point in the Dalitz plot can have a different
strong phase $\delta$ thus complicating the interpretation via
Eqn.~\ref{eqn:ws}; nonetheless, a significant $t^2e^{-t}$ component
in the proper time distribution would be evidence for \DZ\DZBAR\ 
mixing.


 Finally, \cleo\ has searched for evidence of direct \CPV\ in the 
Cabibbo-suppressed processes $\DZ\to K^+K^-$ and $\DZ\to\pi^+\pi^-$.
The initial \DZ\ or \DZBAR\ is tagged by the $\pi^\pm_{\rm slow}$
from ${\rm D}^{*\pm}$ decay and the \DZ\ and \DZBAR\ rates are
extracted from a fit to the $Q$-distribution with the signal shape
taken from Cabibbo-favored $\DZ\to K^-\pi^+$ decays in data and
the background shape taken from simulation. No reconstruction- or
detector-induced asymmetry in the $\pi^\pm_{\rm slow}$ selection, 
$\ACP =  (+0.12\pm 0.36)\%$, is observed as determined from $\KS\to\pi^+\pi^-$
decays.
No significant \CPV\ is observed
{$\ACP(KK) = (0.04  \pm 2.18  \pm 0.84)\% $}  and 
{$\ACP(\pi\pi) = (1.94  \pm 3.22  \pm 0.84)\% $} (preliminary).
The systematic uncertainty from the background shape uncertainty
is estimated to be 0.69\% and the uncertainty due to $\pi^\pm_{\rm slow}$ 
selection is taken as  0.48\%.

In summary no evidence for \CPV\ has been observed by \cleo\ in 
beauty and charm meson decays with a precision of 4\%-25\% (beauty)
and 2-3\% (charm) which is dominated by the statistical
uncertainty. Integrated luminosities approximately 100 times that
accumulated by \cleo\ will be needed to attain a statistical precision
comparable to the magnitude of direct \CPV\ expected in the SM for
beauty and charm decay of ${\cal O}(1\%)$ and  ${\cal O}(0.1\%)$,
respectively. The promising turn-on of the B-factories~\cite{mh,ys} 
indicates that such data samples may be accumulated in approximately
five years or less. Such measurements will then need to confront the 
potentially difficult task of measuring sub-percent 
detector- and reconstruction-induced asymmetries.

I would like to thank the conference organizers for an enjoyable and
informative meeting in beautiful Ferrara. Thanks also to Jesse
Ernst for comments on this contribution.

\end{document}